\documentclass[12pt]{article}
\usepackage{jeosman}
\usepackage{epsfig}
\usepackage{amsmath}
\usepackage{amssymb}
\usepackage{mathrsfs}

\newcommand{\calA}{\mathscr A}
\newcommand{\calR}{\mathscr R}
\newcommand{\calP}{\mathscr P}

\newcommand{\rrr}{{\boldsymbol r}}

\newcommand{\HHH}{\boldsymbol H}
\newcommand{\BBB}{\boldsymbol B}

\newcommand{\nablabf}{\boldsymbol \nabla}
\newcommand{\Laplace}{{\boldsymbol \nabla}^2}

\begin{document}

\title{Mesoscopic magnetism in dielectric photonic crystal meta materials: topology and inhomogeneous broadening}

\author{Niels Asger Mortensen and Sanshui Xiao}

\address{MIC -- Department of Micro and Nanotechnology, NanoDTU,\\
Technical University of Denmark, Bldg.~345 east, DK-2800 Kongens
Lyngby, Denmark}

\email{nam@mic.dtu.dk}

\author{Didier Felbacq}
\address{GES UMR-CNRS 5650, Universit\'e de Montpellier II, B\^at.
21, CC074, Place E. Bataillon, 34095 Montpellier Cedex 05, France}


\begin{abstract}
We consider meta materials made from a two-dimensional dielectric
rod-type photonic crystal. The magnetic response is studied within
the recently developed homogenization theory and we in particular
study the effects of topology and inhomogeneous broadening. While
topology itself mainly affects the Mie resonance frequency we find
that the dispersion in the topological radius $\calR$ of the
dielectric rods may lead to significant inhomogeneous broadening
and suppression of the negative-$\mu$ phenomena for
$\delta\calR/\calR_0\gg \varepsilon''/\varepsilon'$, with
$\varepsilon=\varepsilon'+i\varepsilon''$ being the relative
dielectric function of the rods.
\end{abstract}

\maketitle

\section{Introduction}

\begin{figure}[t!]
\begin{center}
\epsfig{file=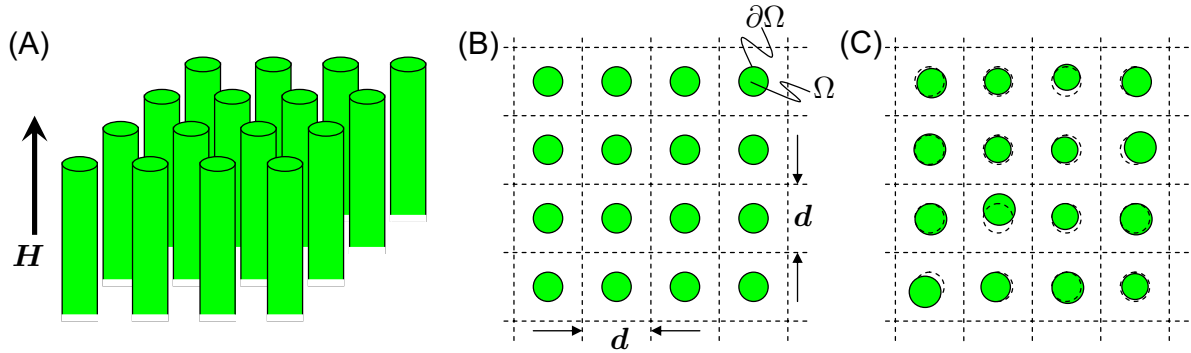, width=\columnwidth,clip}
\end{center}
\caption{(A) Photonic crystal consisting of parallel high-index
dielectric rods. We consider the p-polarized case with the
magnetic field parallel to the rods. (B) Top view of the photonic
crystal with rods arranged in a periodic lattice with pitch $d$.
(C) Top view of weakly disordered/random photonic crystal.}
\label{fig1}
\end{figure}

\begin{figure}[t!]
\begin{center}
\epsfig{file=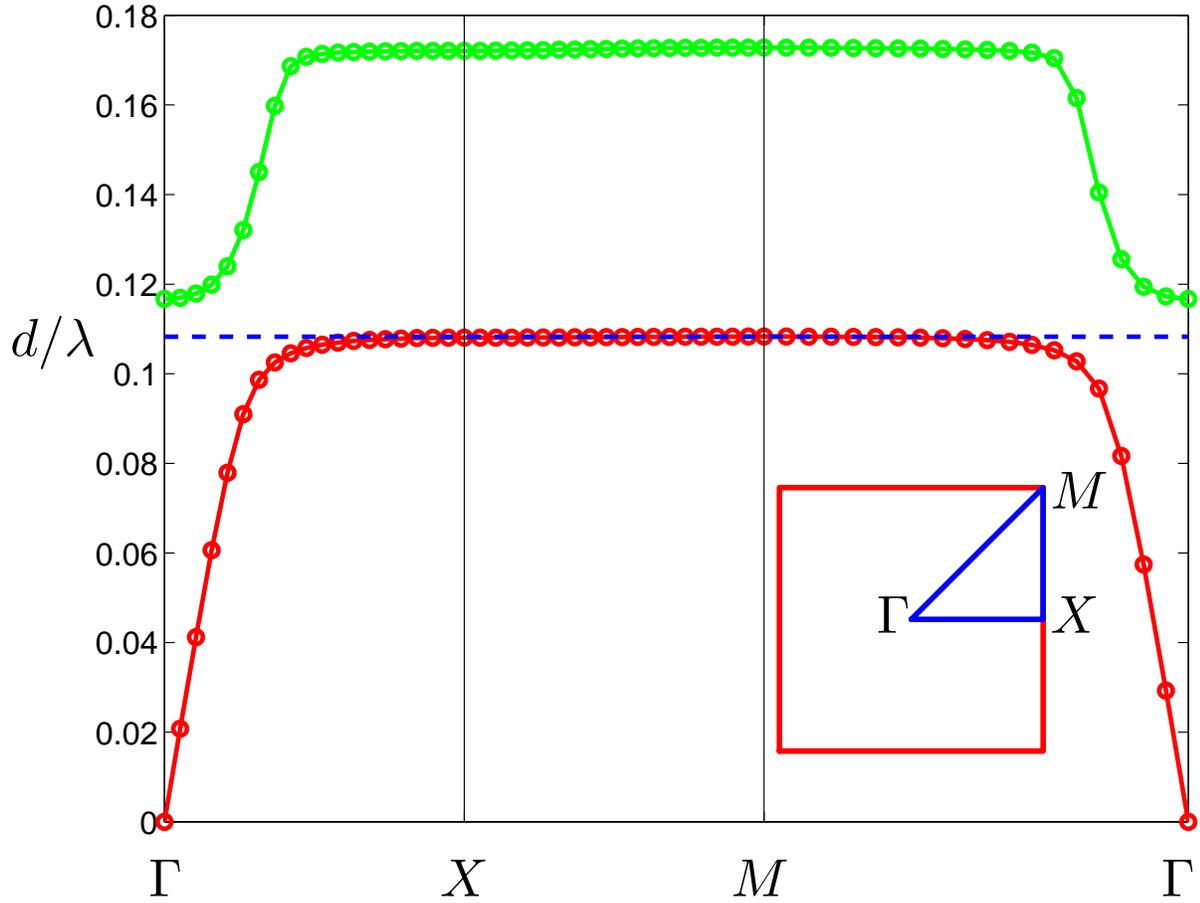, width=\columnwidth,clip}
\end{center}
\caption{Plane-wave simulations of photonic band structure for a
$d\times d$ square lattice of rods of circular cross section with
$\varepsilon=200$, see Fig.~\ref{fig1}(B). The air-filling
fraction $f\simeq 80.37\%$ corresponding to a rod of radius
$R=\calR=d/4$. The dashed line shows the fundamental Mie resonance
of an isolated dielectric rod, Eq.~(\ref{eq:lambda_resonance}).}
\label{fig2}
\end{figure}

The development of meta materials has led to renewed interest in
the original work by Veselago on negative
refraction~\cite{Veselago:1968} and the field of meta materials is
in rapid growth. For more details we refer to recent reviews
~\cite{Pendry:2004b,Smith:2004a,Soukoulis:2006a} as well as to
recent special issues~\cite{Pendry:2003c,Lakhtakia:2005a}. In this
work we consider a photonic crystal composed of high-index
dielectric rods arranged in a two-dimensional lattice, see
Fig.~\ref{fig1}(A). For temporal harmonic modes the $\HHH$ field
is governed by~\cite{Jackson:1962,Joannopoulos:1995}
\begin{equation}\label{eq:waveequation}
\nablabf\times\frac{1}{\varepsilon(\rrr)}\nablabf\times\HHH(\rrr)=\mu(\rrr)\frac{\omega^2}{c^2}\HHH(\rrr)
\end{equation}
where $\varepsilon$ is the relative dielectric permittivity, $\mu$
is the relative magnetic permeability, $c$ is the speed of light in
vacuum, and $\omega=c k$ is the angular frequency with
$k=2\pi/\lambda$ being the free-space wave vector. For Nature's own
dielectric materials we locally have, almost by definition, that
$\mu(\rrr)\equiv 1$ so that everywhere in space
$\BBB(\rrr)=\mu(\rrr)
\HHH(\rrr)\equiv\HHH(\rrr)$~\cite{Joannopoulos:1995}. However, for
artificial composite dielectric structures in the meta material
limit, where the free-space wavelength $\lambda$ is much longer than
the characteristic length scale (periodicity) $d$ of the underlying
(quasi) periodic structure, one may for the slowly varying
homogenized fields (envelopes) have an effective permeability
$\mu_{\rm eff}$ which is constant in space and which at certain
frequencies deviates remarkably from the microscopic permeability,
i.e. $\mu_{\rm eff}\neq \mu$ and in particular it may become
negative. The same apparent paradox is also possible for the
dielectric response. As an example Fig.~\ref{fig2} shows a
plane-wave simulation~\cite{Johnson:2001} of
Eq.~(\ref{eq:waveequation}) for the structure in Fig.~\ref{fig1}A
with $\varepsilon=200$ and $\mu=1$ for the rods and
$\varepsilon=\mu=1$ for the background air region. For these inverse
photonic crystal structures, the band structure and the photonic
band-gap formation can be understood in terms of Mie
resonances~\cite{Sigalas:1996,Lidorikis:2000}. Clearly, the band
structure reveals a strong hybridization at the frequency
corresponding to the first Mie resonance of the rods as indicated by
the dashed line. It is this resonance behaviour which make the rods
behave as artificial magnetic atoms with the possibility for a
negative-$\mu$ response.

In this work we extract the effective permeability by applying the
recently derived mesoscopic homogenization theory for p-polarized
fields by Felbacq and
Bouchitt\'e~\cite{Felbacq:2005b,Felbacq:2005,Felbacq:2005a}. We in
detail study the effective permeability $\mu_{\rm eff}$ and the
effect of the topology of the rods as well as the influence of
spatial disorder/randomness and dispersion in the rod topology,
compare panels (B) and (C) in Fig.~\ref{fig1}. For rods with a
more arbitrary cross section $\Omega$ we define a topological
radius $\calR$ by
\begin{equation}\label{eq:Rtop}
{\calR}= 2{\calA}/{\calP}
\end{equation}
where $\calA$ is the area of $\Omega$ and $\calP$ is the perimeter
of its surface $\partial\Omega$, see Fig.~\ref{fig1}(B). It is
easily verified that a circular cross section of radius $R$ gives
$\calR=R$. More generally, the topological radius $\calR$ captures
the qualitative resonance behaviour of the effective permeability
as we shall see in the following.

Our results for more arbitrary topologies rely on approximative
results for the spectral problem of the Helmholtz equation as
employed recently by one of us in the context of diffusion
dynamics, fluid flow, and ion transport in microfluidic
problems~\cite{mortensen:2006a,mortensen:2006b,mortensen:2006c}.

\section{Multi-scale expansion theory}
We consider the problem of p-polarized fields corresponding to a
magnetic field aligned parallel to the rods, see
Fig.~\ref{fig1}(A). In the limit where $\lambda\gg d$ the problem
posed by Maxwell's equations can be treated by a multi-scale
expansion approach leading to an effective magnetic permeability
given by~\cite{Felbacq:2005}
\begin{align}\label{eq:mumultiscale}
\mu_{\rm eff}&= f\times \mu_{\rm air}+(1-f)\times\mu_{\Omega}\nonumber\\
&= f+(1-f)\frac{\big<1\big|m\big> }{\big<1\big|1\big>}
\end{align}
where $\calA=\big<1\big|1\big>$ is the area of $\Omega$, $f$ is
the air-filling fraction, and $\big|m\big>$ is governed by
\begin{subequations}\label{eq:mueff}
\begin{align}
\Laplace \big|m\big> + k^2 \varepsilon \big|m\big>=0,&\quad \rrr\in\Omega\\
\big|m\big>=1,&\quad \rrr\in\partial\Omega.
\end{align}
\end{subequations}
Above, we have used the Dirac \emph{bra-ket} notation and inner
products are defined as
\begin{equation}
\big<h\big|g\big>=\int_{\Omega}dr\,hg,
\end{equation}
where $h(\rrr)$ and $g(\rrr)$ are functions on $\Omega$. Similarly,
$\big|m\big>$, or $m(\rrr)$, in Eqs. (3) and (4) corresponds to the
magnetic field component along the rods and $\big|1\big>$ denotes
the unity function.

In the following we assume a quadratic unit cell with area $d^2$,
see Fig.~\ref{fig1}(B), so that $f=1-\calA/d^2$. As in
Refs.~\cite{Felbacq:2005b,Felbacq:2005,Felbacq:2005a} we solve the
problem in a complete basis defined by the Helmholtz eigenvalue
equation
\begin{subequations}
\begin{align}
\Laplace \big|\psi_n\big> + \kappa_n^2 \big|\psi_n\big>&=0,\quad \rrr\in\Omega,\\
\big|\psi_n\big>&=0, \quad \rrr\in\partial\Omega,
\end{align}
\end{subequations}
with the basis functions being orthonormal, i.e.
$\big<\psi_m\big|\psi_n\big>=\delta_{nm}$. Obviously, the Ansatz
\begin{equation}
\big|m\big>= \big|1\big>+\sum_n \alpha_n \big|\psi_n\big>
\end{equation}
fulfills the boundary condition and the expansion coefficients are
readily determined by standard \emph{bra-ket} manipulations. In
this way we arrive at the following general expression for the
effective permeability~\cite{Felbacq:2005b}
\begin{equation}\label{eq:muexpansion}
\mu_{\rm eff}= 1+\sum_n
\frac{-k^2\varepsilon}{k^2\varepsilon-\kappa_n^2}
\frac{\left|\big<1\big|\psi_n\big>\right|^2}{d^2}.
\end{equation}
The strength of this result is obviously the connection of
$\mu_{\rm eff}$ to the spectrum and eigenstates of the Helmholtz
equation which have been studied in great detail in various
contexts of physics, including membrane dynamics, the acoustics of
drums, the single-particle eigenstates of 2D quantum dots, not to
forget quantized conductance of quantum wires. Finally, the
results of course in turn provides the direct link between
$\mu_{\rm eff}$ and the Mie resonances.

\section{Approximate properties of the spectral problem}

In the following we utilize approximate results for the spectral
problem and its dependence on geometry. We start by introducing
the effective area $\calA_n=\left|\big<1\big|\psi_n\big>\right|^2$
so that Eq.~(\ref{eq:muexpansion}) is rewritten as
\begin{align}
\mu_{\rm eff}&= 1+(1-f)\sum_n
\frac{-k^2\varepsilon}{k^2\varepsilon-\kappa_n^2}\frac{\calA_n}{\calA}\nonumber\\
&\simeq 1+(1-f)
\frac{-k^2\varepsilon}{k^2\varepsilon-\kappa_1^2}\frac{\calA_1}{\calA}\label{eq:muexpansion_truncation}
\end{align}
where we in the second line have truncated the rapidly converging
sum making the approximation valid in the vicinity of the main
resonance.

In Table~\ref{tab1} we have listed the first eigenvalue $\kappa_1$
and the corresponding effective area $\calA_1$ for a number of
different geometries. For the numerical results we have employed a
finite-element method with an adaptive mesh algorithm (Comsol
multiphysics) ensuring fast convergence of both eigenvalues and
eigenstates. Inspecting the column for the eigenvalue $\kappa_1^2$
in Table~\ref{tab1} we note that the numbers do not in general
change much from geometry to geometry, despite the variation in
topology. The small spread originates in the normalization by the
topological radius $\calR$, see Eq.~(\ref{eq:Rtop}). For the
effective area $\calA_1$ we likewise observe a modest variation.
It is worthwhile emphasizing the results for a circular cross
section which become good approximations for also other relatively
compact topologies. In the following we thus use the approximation
that $\kappa_1\sim \chi_{00}/\calR$ and $\calA_1/\calA\sim
4/\chi_{00}^2$ where $\chi_{00}\simeq 2.405$ is the first zero of
the Bessel function, i.e. $J_0(\chi_{00})=0$. With these
approximations we finally arrive at
\begin{align}
\mu_{\rm eff}&\simeq 1+(1-f)\frac{4}{\chi_{00}^2}
\frac{-(k\calR)^2\varepsilon}{(k\calR)^2\varepsilon-\chi_{00}^2}\nonumber\\
&= 1+(1-f)\frac{4}{\chi_{00}^2}
\frac{-\varepsilon}{\varepsilon-\left(\frac{\chi_{00}}{2\pi}\right)^2\left(\frac{\lambda}{\calR}\right)^2}\label{eq:mueffapprox}
\end{align}
where all the dependence on geometry is captured by the
topological radius $\calR$.

We may now estimate the resonance wavelength. Suppose that
$\varepsilon=\varepsilon'+i\varepsilon''$, then for weakly
absorbing materials, $\varepsilon''\ll\varepsilon'$, we have the
resonance around
\begin{equation}\label{eq:lambda_resonance}
\lambda^*\simeq \frac{2\pi}{\kappa_1} \sqrt{\varepsilon'}\approx
\frac{2\pi}{\chi_{00}} \sqrt{\varepsilon'}\:\calR
\end{equation}
with a line width of the order
\begin{equation}
\delta\lambda \sim \frac{2\pi}{\chi_{00}}
\frac{\varepsilon''}{\sqrt{\varepsilon'}}\:\calR.
\end{equation}
As expected the line width reflects the damping due to
$\varepsilon''$ in the high-index material. Furthermore, we may
also estimate the magnitude of the effective magnetic response.
Just below the resonance the real part is bounded by
\begin{equation}\label{eq:Remu}
{\rm Re}\big\{ \mu_{\rm eff}\big\}\gtrsim
1-\frac{2}{\chi_{00}^2}(1-f)\frac{\varepsilon'}{\varepsilon''}
\end{equation}
while for the imaginary part
\begin{equation}\label{eq:Immu}
{\rm Im}\big\{ \mu_{\rm eff}\big\}\lesssim
\frac{4}{\chi_{00}^2}(1-f)\frac{\varepsilon'}{\varepsilon''}.
\end{equation}
The above results illustrate how $\varepsilon$, $f$, and $\calR$
influence the magnetic response. Obviously, a negative
permeability requires that $\varepsilon'\gg\varepsilon''$. On the
other hand, some absorption is of course needed to have a finite
line width instead of a delta-function-like response which would
suffer tremendously from the inevitable effects of inhomogeneous
broadening. We shall return to this in
Sec.~\ref{sec:inhomogeneous}.

\section{Comparison to numerical results}

\begin{figure}[t!]
\begin{center}
\epsfig{file=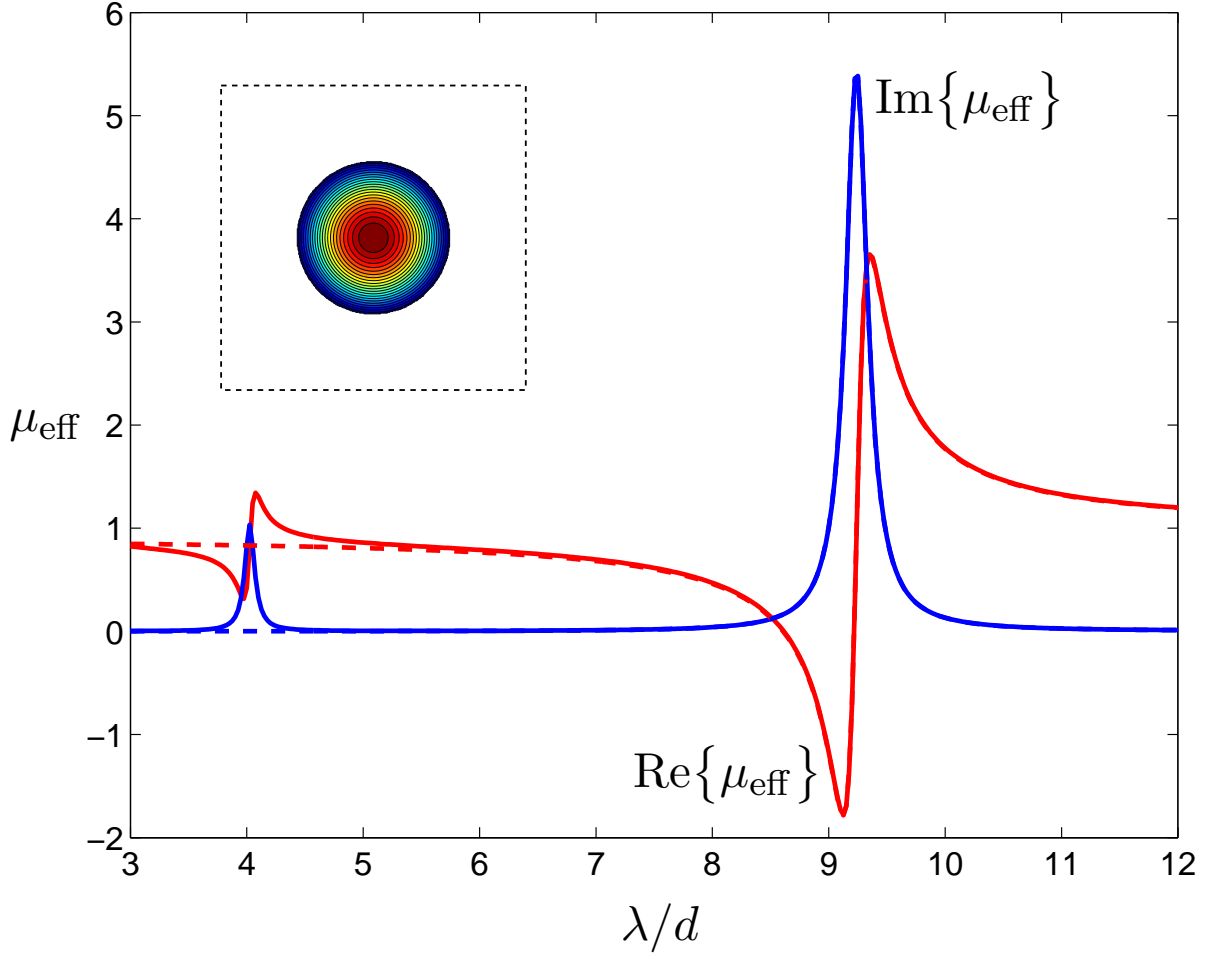, width=\columnwidth,clip}
\end{center}
\caption{Effective permeability $\mu_{\rm eff}$ versus normalized
wavelength $\lambda/d$ for a rod with $\varepsilon=200+5i$ in a
quadratic $d\times d$ unit cell with an air-filling fraction
$f\simeq 80.37\%$ corresponding to a rod of radius $R=\calR=d/4$.
The inset illustrates the unit cell as well as the function $m$ at
a wavelength above resonance. The solid lines are numerical exact
solutions based on a finite-element (adaptive mesh) solution of
Eq.~(\ref{eq:mueff}) while the dashed lines show the approximate
result in Eq.~(\ref{eq:mueffapprox}).} \label{fig3}
\end{figure}

\begin{figure}[t!]
\begin{center}
\epsfig{file=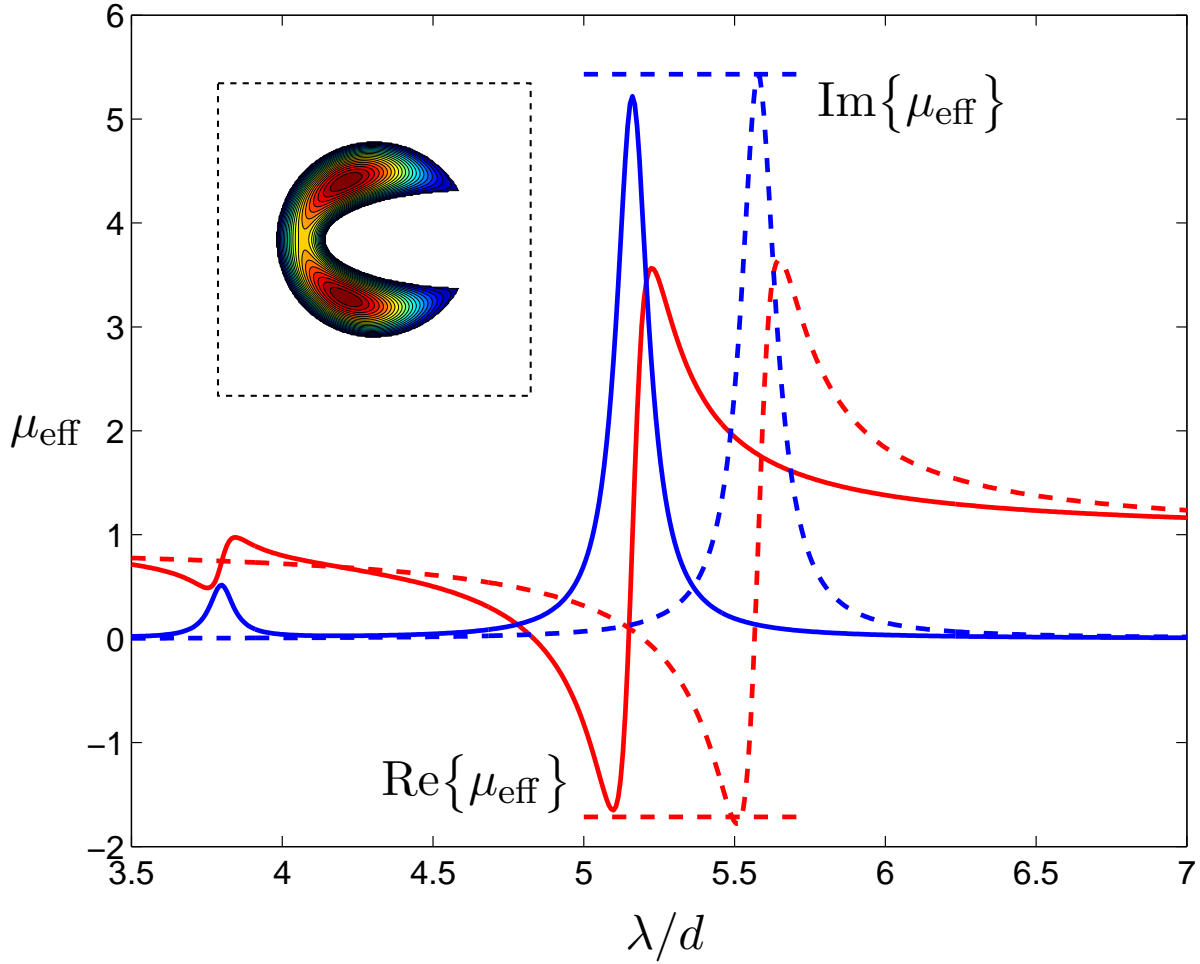, width=\columnwidth,clip}
\end{center}
\caption{Effective permeability $\mu_{\rm eff}$ versus normalized
wavelength $\lambda/d$ for a rod with $\varepsilon=200+5i$ in a
quadratic $d\times d$ unit cell with an air-filling fraction
$f\simeq 80.37\%$. The inset illustrates the unit cell as well as
the function $m$ at a wavelength above resonance. The highly
non-circular cross section has a topological radius of
$\calR\simeq 0.151\times d$. The solid lines are numerical exact
solutions based on a finite-element (adaptive mesh) solution of
Eq.~(\ref{eq:mueff}) while the dashed lines show the approximate
result in Eq.~(\ref{eq:mueffapprox}). The horizontal dashed lines
show the limits in Eqs.~(\ref{eq:Remu}) and (\ref{eq:Immu}). }
\label{fig4}
\end{figure}

The validity of the assumptions underlying
Eq.~(\ref{eq:mumultiscale}) has been tested in
Refs.~\cite{Felbacq:2005b,Felbacq:2005,Felbacq:2005a} by comparing
simulated transmission spectra of the real photonic crystal to
spectra based on the corresponding homogenized meta material. In
this work we thus assume Eq.~(\ref{eq:mumultiscale}) valid and
focus on checking the quality of the approximate result in
Eq.~(\ref{eq:mueffapprox}). We do this by comparing the outcome of
Eq.~(\ref{eq:mueffapprox}) to numerical solutions of the original
problem posed by Eq.~(\ref{eq:mumultiscale}). We use a
finite-element method with an adaptive mesh algorithm (Comsol
multiphysics) for this purpose. For the dielectric function we use
$\varepsilon=200+5i$ as in
Ref.~\cite{Felbacq:2005b,Felbacq:2005,Felbacq:2005a}.
Fig.~\ref{fig2} illustrates the corresponding photonic band
structure. Clearly, there is a strong hybridization at the
frequency corresponding to the isolated Mie resonance of the rod,
see Eq.~(\ref{eq:lambda_resonance}). It is this avoided crossing
which causes strong resonance in the effective permeability that
is observed in Fig.~\ref{fig3}. The approximate result in
Eq.~(\ref{eq:mueffapprox}) works excellently near the main
resonance for rods with circular cross section as seen in
Fig.~\ref{fig3} while the approximation of course doesn't capture
high-order modes due to the truncation of the summation in
Eq.~(\ref{eq:muexpansion_truncation}). In the following we go a
step further and test the result for highly non-circular
geometries. Fig.~\ref{fig4} illustrates results for one such
example. As seen there is a good qualitative agreement and since
the geometry is highly non-circular, see inset, the quantitative
agreement is perhaps better than expected. Note how the change in
topological radius $\calR$ scales the resonance wavelength
$\lambda^*$ and the line width $\delta\lambda$ compared to the
circular case, Fig.~\ref{fig3}, in accordance with
Eq.~(\ref{eq:lambda_resonance}) while the conservation of the
air-filling fraction $f$ leads to the same magnitude of the
response in accordance with Eqs.~(\ref{eq:Remu}) and
(\ref{eq:Immu}) which depend only on $f$, but not on $\calR$.

\section{Inhomogeneous broadening}\label{sec:inhomogeneous}

\begin{figure}[t!]
\begin{center}
\epsfig{file=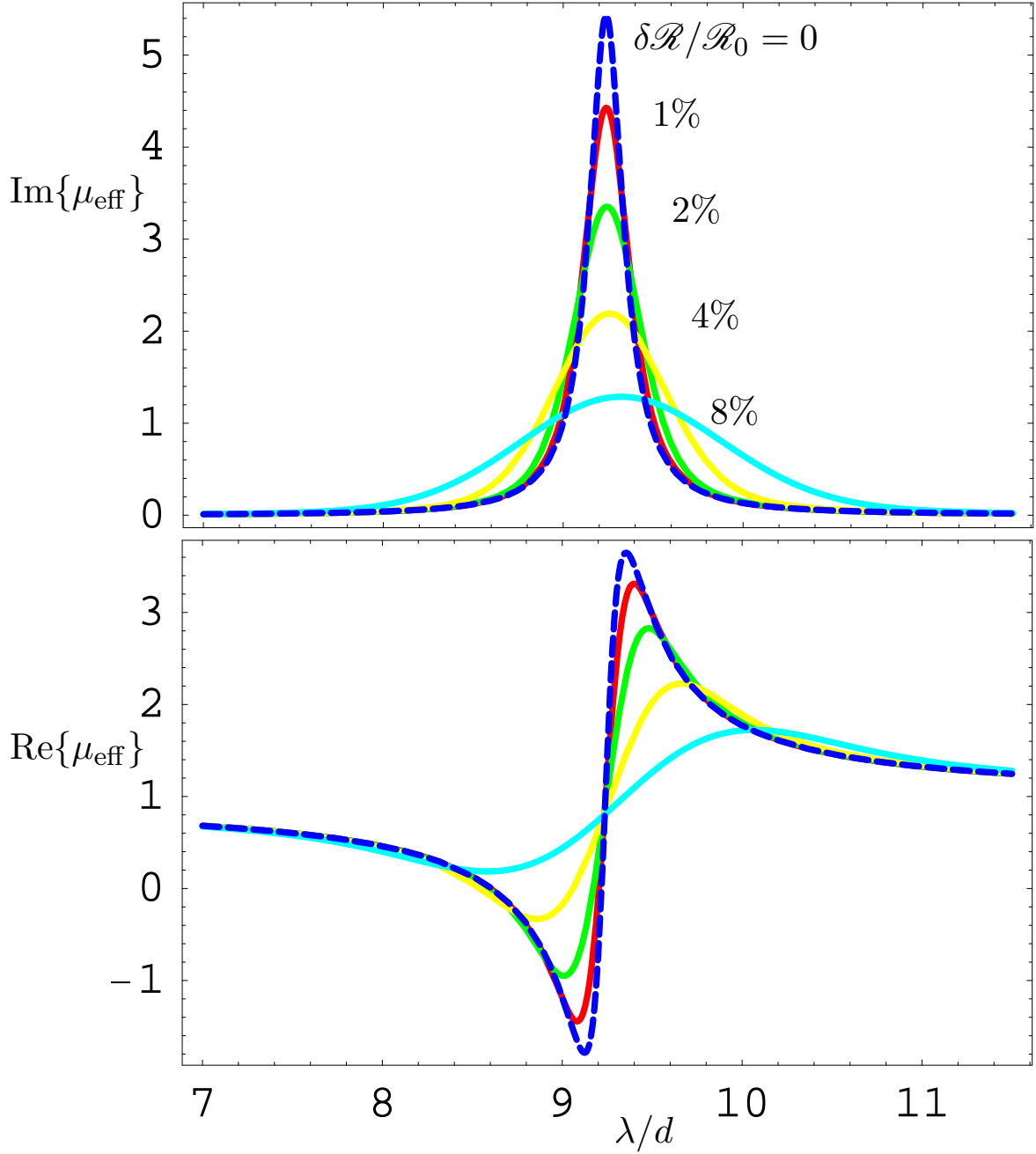, width=\columnwidth,clip}
\end{center}
\caption{Inhomogeneous broadening of the effective permeability
$\mu_{\rm eff}$ for rods with a circular cross section in a
quadratic $d\times d$ unit cell with an average air-filling
fraction $f_0\simeq 80.37\%$ corresponding to rods with an average
radius $R_0=\calR_0=d/4$. The dashed curves show the intrinsic
line width corresponding to the results in Fig.~\ref{fig3}. }
\label{fig5}
\end{figure}

The present results suggest that shape plays only a minor role and
mainly affects the particular value of the resonance frequency
(which is proportional to the topological radius $\calR$) while
the air-filling fraction $f$ and the dielectric function
$\varepsilon=\varepsilon'+i\varepsilon''$ tend to govern the
magnitude of the effective magnetic response. This suggests that
homogenized materials will not be strongly influenced by spatial
disorder/randomness conserving the average air-filling fraction
while shape dispersion (dispersion in the topological radius,
possibly also influencing the average air-filling fraction) will
give rise to a smearing of the resonance phenomena. Suppose that
for a particular topological radius $\calR_0$ the resonance
wavelength is $\lambda_0^* = \alpha \calR_0$, see
Eq.~(\ref{eq:lambda_resonance}), and that the width of the
resonance is $\delta\lambda_0$. Assuming a Gaussian distribution
of $\calR$,
\begin{equation}
P(\calR)\propto\exp\left[-\left(\frac{\calR-\calR_0}{\delta\calR}\right)^2\right],
\end{equation}
and neglecting the dependence of the intrinsic line width
$\delta\lambda_0$ on $\calR$ we get an inhomogeneous suppression
of the resonance magnitude by a factor $\Pi$,
\begin{equation}
\Pi=\frac{1}{\sqrt{1+\alpha^2\left(\frac{
\delta\calR}{\delta\lambda_0}\right)^2}},
\end{equation}
and an inhomogeneous broadening of the resonance by a factor
$\Pi^{-1}$,
\begin{equation}
\delta\lambda\sim \Pi^{-1}\delta\lambda_0.
\end{equation}
Thus, inhomogeneous broadening will be negligible if
$\alpha\left(\frac{ \delta\calR}{\delta\lambda_0}\right)\ll 1$
corresponding to
\begin{equation}\label{eq:criterion}
\frac{\delta\calR}{\calR_0}\ll \frac{\varepsilon''}{\varepsilon'}.
\end{equation}
Fig.~\ref{fig5} illustrates these results in the case of rods of
circular cross section as in Fig.~\ref{fig3}. The results are
obtained from a numerical evaluation of $\int_0^\infty d\calR\,
\mu_{\rm eff}(\calR)P(\calR)$ with $\mu_{\rm eff}$ given by
Eq.~(\ref{eq:mueffapprox}). As also suggest by
Eq.~(\ref{eq:criterion}) a size dispersion with a relative
variation comparable to $\varepsilon''/\varepsilon'$ preserves the
negative-$\mu$ phenomena while larger variations will result in a
positive $\mu_{\rm eff}$.

\section{Conclusion}

We have studied the magnetic response of rod-type dielectric meta
materials within the recently developed homogenization theory
\cite{Felbacq:2005b} and we have in particular investigated the
effects of topology and inhomogeneous broadening. We predict that
topology itself mainly affects the Mie resonance frequency while
the resonance shape mainly depends on the air-filling fraction and
the ratio $\varepsilon'/\varepsilon''$ with
$\varepsilon=\varepsilon'+i\varepsilon''$ being the relative
dielectric function of the rods. Furthermore, we argue that
randomness/disorder conserving the air-filling fraction will have
a limited influence on the negative-$\mu$ phenomena while
dispersion in the topological radius $\calR$ of the rods may lead
to significant inhomogeneous broadening and suppression of the
phenomena.

\section*{Acknowledgments}

This work is financially supported by the \emph{Danish Council for
Strategic Research} through the \emph{Strategic Program for Young
Researchers} (grant no: 2117-05-0037).

\newpage


\newpage

\begin{table}[t!]
\begin{center}
\begin{tabular}{lcccc}
& $\left(\kappa_1 {\calR}\right)^2$ & ${\calA}_1/{\calA}$
\\\hline
circle & $\chi_{00}^2\simeq 5.78$$^{a}$ & $4/\chi_{00}^2\simeq 0.69$$^{a}$ \\
quarter-circle  & 5.08$^b$ & 0.65$^b$  \\
half-circle     & 5.52$^b$ & 0.64$^b$ \\
ellipse(1:2)    & 6.00$^b$ & 0.67$^b$\\
ellipse(1:3)    & 6.16$^b$ & 0.62$^b$\\
ellipse(1:4)    & 6.28$^b$ & 0.58$^b$\\\hline
triangle(1:1:1) & $4\pi^2/9\simeq 4.39$$^c$& $6/\pi^2\simeq 0.61$$^c$ \\
triangle(1:1:$\sqrt{2}$) & $\frac{5\pi^2}
  {( 2 + \sqrt{2})^2}\simeq 4.23$$^a$ &$512/9\pi^4\simeq 0.58$$^a$
\\\hline
square(1:1) & $\pi^2/2\simeq 4.93$$^a$ & $64/\pi^4\simeq 0.66$$^a$\\
rectangle(1:2) & $5\pi^2/9\simeq 5.48$$^a$& $64/\pi^4\simeq 0.66$$^a$\\
rectangle(1:3) & $ 5\pi^2/8\simeq 6.17$$^a$ & $64/\pi^4\simeq 0.66$$^a$\\
rectangle(1:4) & $17\pi^2/25\simeq 6.71$$^a$ &$64/\pi^4\simeq
0.66$$^a$\\\hline
pentagon &  5.20$^b$ & 0.67$^b$\\\hline
hexagon& 5.36$^b$ & 0.68$^b$\\\hline
\end{tabular}
\caption{Central dimensionless parameters for different
geometries. $^a$See e.g.~\cite{Morse:1953} for the eigenmodes and
eigenspectrum. $^b$Data obtained by finite-element simulations.
$^c$See e.g.~\cite{Brack:1997} for the eigenmodes and
eigenspectrum. } \label{tab1}
\end{center}
\end{table}

\end{document}